\documentclass[journal]{IEEEtran}
\usepackage{orcidlink}
\usepackage{url}
\usepackage{indentfirst}
\usepackage{cite}
\usepackage{booktabs}
\usepackage{amsmath,amssymb}
\usepackage{graphicx}
\usepackage{subfigure}
\usepackage{multirow}

\title{Towards a Multi-class Classification of Upper Limb Movements}

\author{Hao Jia\orcidlink{0000-0003-1356-7463}, 
        Fan Feng\orcidlink{0000-0002-3913-591X},
        Cesar F. Caiafa\orcidlink{0000-0001-5437-6095}, 
        Feng Duan~\IEEEmembership{Member,~IEEE}\orcidlink{0000-0002-2179-2460}, 
        Yu Zhang\orcidlink{0000-0003-4087-6544}, 
        Zhe Sun\orcidlink{0000-0002-6531-0769}, 
        Jordi Sol{\'e}-Casals\orcidlink{0000-0002-6534-1979}
\thanks{This work was carried out as part of the doctoral programme in Experimental Science and Technology at the University of Vic - Central University of Catalonia. This research was supported by the National Key R\&D Program of China (No. 2017YFE0129700), the National Natural Science Foundation of China (Key Program) (No. 11932013), the National Natural Science Foundation of China (No. 61673224), the Tianjin Natural Science Foundation for Distinguished Young Scholars (No. 18JCJQJC46100), the Tianjin Science and Technology Plan Project (No. 18ZXJMTG00260) and the Ministerio de Economia y Competitividad of Spain (No. TEC2016-77791-C4-2-R). J.S-C. work is also based upon work from COST Action CA18106, supported by COST (European Cooperation in Science and Technology). C.F.C work is partially supported by grants UBACyT 20020190200305BA, UBACyT 20020170100192BA, and MINCyT PICT 3208 2017 (Argentina). \textit{(Corresponding author: Feng Duan, Zhe Sun, Jordi Sol{\'e}-Casals.)}}
\thanks{Hao Jia is currently the Ph.D. student in the Data and Signal Processing Research Group, University of Vic-Central University of Catalonia, Vic, Catalonia.}
\thanks{Fan Feng is currently working in the College of Artificial Intelligence, Nankai University, Tianjin, China.}
\thanks{Cesar F. Caiafa is currently working in the Instituto Argentino de Radioastronom\'{i}a, CONICET CCT La Plata/CIC-PBA/UNLP, V. Elisa, Argentina, and Visiting Researcher in the College of Artificial Intelligence, Nankai University, Tianjin, China.}
\thanks{Feng Duan is currently working in the College of Artificial Intelligence, Nankai University, Tianjin, China (email: duanf@nankai.edu.cn).}
\thanks{Yu Zhang is currently working in the Department of Bioengineering, Lehigh University,
Bethlehem, PA 18015, USA.}
\thanks{Zhe Sun is currently working in the Computational Engineering Applications Unit, Head Office for Information Systems and Cybersecurity, RIKEN, Saitama, Japan (email: zhe.sun.vk@riken.jp).}
\thanks{Jordi Sol{\'e}-Casals is currently working in the Data and Signal Processing Research Group, University of Vic - Central University of Catalonia, Vic, Catalonia, and Visiting Researcher in the Department of Psychiatry, University of Cambridge, United Kingdom and in the College of Artificial Intelligence, Nankai University, Tianjin, China (email: jordi.sole@uvic.cat).}}
\begin{document}

\maketitle
\begin{abstract}
Non-invasive brain-computer interfaces can help humans to control external devices. Previous studies on the classification of limb movements have focused on the classification of left/right limbs; however, the classification of different types of upper limb movements has often been ignored despite that it provides more active-evoked control commands in the brain-computer interface. Nevertheless, no machine learning method can be used as the baseline method in the multi-class classification of limb movements.

This work focuses on the multi-class classification of upper limb movements and proposes the multi-class filter bank task-related component analysis (mFBTRCA) method, which consists of three steps: spatial filtering, similarity measuring and filter bank selection. The spatial filter, namely the task-related component analysis, is first used to remove noise from EEG signals. The canonical correlation measures the similarity of the spatial-filtered signals and is used for feature extraction. The correlation features are extracted from multiple low-frequency filter banks. The minimum-redundancy maximum-relevance method selects the essential features from all the correlation features, and finally, the support vector machine is used to classify the selected features.

The proposed method compared against previously used models is evaluated using two datasets. In these, mFBTRCA achieved a classification accuracy of 0.4022$\pm$0.0709 (7 classes) and 0.4032$\pm$0.0739 (5 classes), respectively, which improves on the best accuracies achieved using the compared methods (0.3603$\pm$0.0640 and 0.3159$\pm$0.0736, respectively). The proposed method is expected to provide more control commands in the applications of non-invasive brain-computer interfaces.
\end{abstract}

\begin{IEEEkeywords}
Brain-computer Interface,
Electroencephalogram,
Movement-related Cortical Potential,
Upper Limb Movement,
Pattern Recognition.
\end{IEEEkeywords}

\section{Introduction}
\label{sec:intro}
\IEEEPARstart{A}{non-invasive} brain-computer interface is a framework that bridges the gap between human brains and external computers \cite{schalk_practical_2010,lopez-larraz_brain-machine_2018}. Electroencephalogram (EEG) signals are multi-channel time series acquired from human scalps using a brain-computer interface system. EEG signals are usually used to generate multiple commands by classifying certain tasks, such as left and right limb movements or multiple visual stimuli. These commands can be used to control robots or other external devices \cite{liang_exploring_2021}. 

In current research on brain-computer interfaces, brain activities such as motor imagery and steady-state visual evoked potentials are frequently used in human-robot interactions. In motor imagery, the commands are generated by classifying movements of the left/right hand, a foot or the tongue. These active-evoked commands are controlled by human intent. In steady-state visual evoked potentials, the number of commands depends on the number of visual stimuli, and hence there are more control commands. However, the steady-state visual evoked potential is evoked by external visual stimuli. When there are no external visual stimuli, the subjects are unable to generate these control commands intentionally. Thus, the passive-evoked commands limit the application of steady-state visual evoked potentials. Movement-related cortical potential (MRCP) is a brain activity related to limb movement \cite{shakeel_review_2015}. However, current approaches have only focused on the binary classification between the limb's resting and movement states.

Most brain-computer interface studies focus on improving existing classification tasks in motor imagery and steady-state visual evoked potentials \cite{Liu_2021,zhang_deep_2020,zhang_temporally_2019,jiao2018novel,nakanishi2018enhancing,liu_improving_2021}. However, the exploration of more user-friendly brain activities has been ignored. Limb movements are active-evoked and are controlled by the intent of the subject. The classification of multiple upper limb movements not only is more friendly to users than visual stimuli, but also has more commands if combined with left and right limb classification. However, there are very few methods for the multi-class classification of limb movements such as elbow flexion and pronation of the same limb \cite{ibanez_predictive_2015,jochumsen_detecting_2016,robinson_noninvasive_2016}. This work focuses on the multi-class classification of limb movements from MRCP signals, and proposes the multi-class filter bank task-related component analysis (mFBTRCA) method to solve the classification problem. A list of acronyms used in this work is included in Table \ref{tab:0} to support the reading of the manuscript.

The structure of this work is as follows. In Section \ref{sec:related}, a review is presented of related works on the classification of limb movements. In Section \ref{sec:method}, the dataset description and details on how the dataset is pre-processed are given. This section also includes a description on the structure of the mFBTRCA method. In Section \ref{sec:result}, the performance of mFBTRCA is evaluated in the binary classification cases. The proposed method is also compared against other multi-class classification benchmark methods in the multi-class cases. In Section \ref{sec:discuss}, a discussion is given on how the mFBTRCA method uses the information from the MRCP signals, and the bottleneck of mFBTRCA in the multi-class limb movement classifications is also touched upon. Finally, conclusions are given in Section \ref{sec:conclude}.

\begin{table}[htbp]
    \centering
    \caption{Acronyms and their Corresponding Full Names}
    \label{tab:0}
    \begin{tabular}{c|c}
    \toprule
        \textbf{Acronym} & \textbf{Full Name} \\
        \midrule
        \multicolumn{2}{c}{Concept} \\
        \midrule
        EEG & Electroencephalogram \\
        MRCP & Movement-Related Component Analysis \\
        \midrule
        \multicolumn{2}{c}{Limb State} \\
        \midrule
        EF & Elbow Flexion \\
        EE & Elbow Extension \\
        SU & Supination \\
        PR & Pronation \\
        HC & Hand Close \\
        HO & Hand Open \\
        RE & Resting \\
        PG & Palmar Grasp \\
        LG & Lateral Grasp\\
        \midrule
        \multicolumn{2}{c}{Machine Learning Method} \\
        \midrule
        CSP & Common Spatial Pattern \\
        mCSP & multi-class Common Spatial Pattern \\
        FBCSP & Filter Bank Common Spatial Pattern \\
        bSTRCA & binary Standard Task-Related Component Analysis \\
        mSTRCA & multi-class Standard Task-Related Component Analysis \\
        bFBTRCA & binary Filter Bank Task-Related Component Analysis \\
        mFBTRCA & multi-class Filter Bank Task-Related Component Analysis\\
        SPoC & Source Power Comodulation \\
        MDM & Minimum Distance to Mean \\
        LDA & Linear Discriminant Analysis \\
        TSLDA & Tangent Space Linear Discriminant Analysis \\
        \midrule
        \multicolumn{2}{c}{Deep Learning Method} \\
        \midrule
        CNN & Convolutional Neural Network \\
        RNN & Recurrent Neural Network \\
        GNN & Graph Neural Network \\
        LSTM & Long Short-Term Memory \\
        GRU & Gate Recurrent Unit \\
        SCNN & Shallow Convolutional Neural Network \\
        DCNN & Deep Convolutional Neural Network \\
        C-R-CNN & Convolutional-RNN-Convolutional Neural Network\\
        C-L-CNN & Convolutional-LSTM-Convolutional Neural Network\\
        C-G-CNN & Convolutional-GRU-Convolutional Neural Network\\
        GC-G-CNN & Graph Convolutional-GRU-Convolutional Neural Network\\
        \bottomrule
    \end{tabular}
\end{table}

\section{Related Works}
\label{sec:related}
In this section, we will first introduce previous multi-class classification methods related to limb movements, including machine learning and deep learning methods. The binary classification methods based on MRCP signals are then presented, which are also related to limb movements.

Amongst brain activities that are related to limb movements, motor imagery is frequently used in brain-computer interfaces. The multi-class classification algorithms of limb movements have mostly been developed based on motor imagery in previous works. Motor imagery is related to the power change of EEG signals between 8 $Hz$ and 30 $Hz$. When the left/right limb moves, the power of the EEG signals in some channels increases or decreases. Common spatial pattern (CSP) is the basic binary classification algorithm used for motor imagery \cite{blankertz_optimizing_2008}. It searches for the spatial filter that maximizes and minimizes the self-covariance of EEG signals, and it then converts the filtered signals to the logarithm variances. The variances are the features used for classification in CSP. The filter bank CSP computes and combines multiple CSPs in various frequency bands, and selects features from these bands by feature selection methods based on mutual information. The filter bank CSP method has demonstrated its competitive performance in several competitions \cite{kai_keng_ang_filter_2008}. However, the original CSP and filter bank CSP algorithms can only be used in binary classification. The multi-class version of CSP put the common spatial pattern in the framework of information theoretic feature extraction \cite{grosse-wentrup_multiclass_2008}. Therefore, the common spatial pattern method can classify multiple limb movements such as those made by the left/right hand, a foot or the tongue. 
In the CSP algorithm, the spatial filtering and the logarithm covariance features can be regarded as the computation of a Riemann distance in the space of covariance matrices in the context of brain-computer interfaces \cite{barachant_common_2010}. The covariance matrices are used as EEG signal descriptors. In the Riemannian geometry, these matrices are classified directly using the topology of the manifold of symmetric and positive definite matrices. The computation on the Riemannian manifold facilitates to  discern between the multiple classes of limb movements. The minimum distance to mean (MDM) is the straightforward algorithm relying on the Riemannian manifold and the Riemannian distance \cite{barachant_multiclass_2012}. This algorithm regards the covariances of EEG signals as points on the manifolds. The center point of the points belonging to the same class is computed with the Riemannian mean of covariances. The classes of points are predicted by measuring the Riemannian distance between points and the center points. The tangent space linear discriminant analysis classifier is an optimized algorithm based on the Riemannian manifold \cite{barachant_multiclass_2012}. It maps a set of covariance matrices to the Riemannian tangent space, and these matrices are vectorized. A linear discriminant classifier is then used to classify the dimensionally-reduced matrices.

Following the development of deep learning techniques, neural networks have proved to be very useful in EEG signal processing due to their competitive classification performances. During EEG acquisition, signals are sparsely sampled from several electrodes on the scalp, such that EEG signals have low-grade spatial characteristics. The convolutional neural network (CNN) has shown its efficiency in extracting and optimizing spatial features. The shallow CNN (SCNN) and the deep CNN (DCNN) are two CNN architectures used for the end-to-end EEG analysis \cite{schirrmeister_deep_2017}. The DCNN has been shown to perform at least as well in binary classification as the widely used filter bank CSP algorithm. Because both the SCNN and DCNN architectures are based on neural networks, and the number of output neurons can be enlarged as is convenient, both can be used in multi-class classification.
As EEG signals are multi-channel time series, the temporal dynamic processes will also be considered in signal processing. The recurrent neural network (RNN) can use the temporal characteristics of EEG signals, especially when equipped with long short-term memory units \cite{ordonez_deep_2016}. In some studies, CNN and RNN were used simultaneously in EEG signal processing to fuse the spatial and temporal features \cite{wu_learning_2022,zhang_making_2020}.

Neural networks can capture the spatial and temporal characteristics of EEG signals. However, the transformation process going from EEG signals to features is obscure. Deep learning in EEG signal classification has achieved impressive performances. The main reasons are (1) the development of deep learning in pattern recognition, and (2) a deeper understanding of the traditional machine learning based algorithms. In motor imagery, CSP-based algorithms have three main steps carried out during feature extraction. Firstly, the spatial filter of CSP selects the most discriminant channels and optimizes the spatial characteristics. FBCSP then divides EEG signals into several filter banks and applies the CSP algorithm in each band, thus optimizing the features in different filter banks. Finally, multiple CSPs can also be applied to the different sliding time windows of EEG signals, and can then select CSP features of these windows \cite{zhang_temporally_2019}. The EEG signal processing procedure can thus be divided into spatial optimization, filter bank selection and time window selection steps. Different network architectures can be interpreted from the three points when applying deep learning techniques to EEG signals. For example, RNN uses the temporal characteristic \cite{ma_novel_2021,krishna_speech_2019,krishna_speech_2020,krishna_stateart_2020,krishna_advancing_2020} while the graph neural network (GNN) optimizes the spatial characteristic of the EEG signals \cite{georgiadis_using_2019,mathur_graph_2020,xu_adaptive_2019,song_eeg_2020,ghazani_graph_2020,hou_deep_2020,zhang_motor_2020,zhang_deep_2020,zhou_graph_2019}.

However, the machine learning based algorithms cannot fully interpret the information in MRCP signals. Compared to the left/right limb movements in motor imagery, MRCP is related to the motions of the limb. The MRCP signals are located at the low-frequency bands of the EEG signals, namely 0.05$\sim$10 $Hz$. As a result, the noise or task-unrelated components in other bands have to be removed in the signal processing. The grand average MRCP is the approach used to visualize the MRCP signals. It removes the noises by taking the average of the EEG signals of multiple trials belonging to the same class.

Based on the grand average MRCP, Niazi \textit{et al.} proposed a matched filter method to solve the binary classification task \cite{niazi_detection_2011}. This method uses a spatial filter to maximize the MRCP energy and minimize the noise energy by optimizing the signal-to-noise ratio of the signal power. After the spatial filtering, it uses the likelihood ratio to match the relationship between the grand average MRCP and the signals before averaging, thus detecting the movement execution.

The manifold-learning method was introduced to MRCP processing by Xu \textit{et al.} and showed improvements compared to the matched filter method \cite{ren_xu_enhanced_2014}. This method projects the signals into the manifold by using locality-preserving projections. In the manifold space, these multi-channel signals are regarded as points. The grand average MRCP is located at the center of the trials belonging to the same class. These EEG signals are classified in the manifold space using linear discriminant analysis. Lin \textit{et al.} optimized the manifold method proposed by Xu \textit{et al.} by constructing the within-class graph and the between-class graph when projecting EEG signals onto the manifold \cite{lin_discriminative_2016}. A nearest-neighbour classifier is used to measure the distances between the grand average MRCP point and other points, thus predicting their labels. The manifold projection in the two methods and the spatial filtering are in fact used to find a matrix to reduce the dimension of the original EEG signals through matrix multiplication. The procedure of the traditional CSP method consists of spatial filtering and feature extraction. From this perspective, the manifold-based methods lack the step of feature extraction, because the signals are classified by comparing the distances after the manifold projection.

In our previous work, we proposed the binary standard task-related component analysis method (bSTRCA) \cite{duan_decoding_2021}. The bSTRCA follows the processing procedure of spatial filtering and feature extraction. The spatial filter used is the task-related component analysis, and the extracted feature is the canonical correlation coefficient. The bSTRCA method is similar to the matched filter method, as both methods first use spatial filtering to reject the noise in EEG signals and then use a similarity measurement to match the unlabelled EEG signals and the grand average MRCP. However, there are two main differences between the two methods. The first is related to how the spatial filter rejects the noise in the signals. The matched filter method and the bSTRCA method carry out the noise rejection based on the variance and the amplitude of the signals, respectively. MRCP signals are located at the low-frequency band in the frequency domain. In said band, the amplitude of signals mainly reflects the energy change of the signals instead of the variances. The second difference is the role that the similarity measurement plays in the classification. In the matched filter method, the likelihood ratio is the indicator used for classifying the movement and resting states by a threshold criterion. In bSTRCA, correlation coefficients are extracted as features, and a linear discriminant analysis classifier is then used to classify the features. Filter bank selection can further optimize the performance of bSTRCA, and hence the binary filter bank task-related component analysis (bFBTRCA) method was proposed \cite{jia_improve_2022}. However, bFBTRCA is not available for multi-class classification because the framework of bSTRCA was initially designed for binary classification.

In this work, we aim to migrate the structure of the bSTRCA method to the multi-class standard task-related component analysis (mSTRCA). Furthermore, we propose the multi-class filter bank task-related component analysis (mFBTRCA) method by incorporating filter bank selection into mSTRCA. The proposed method can be used in the multi-class classification task of limb movements.

\section{Method}
\label{sec:method}
\subsection{Dataset Description}
\label{ssec:dataset}

Two public EEG datasets (namely datasets I and II) were used to evaluate the performance of the proposed method against the state-of-the-art and baseline methods \cite{ofner_upper_2017,ofner_attempted_2019}. In both datasets, the EEG signals were downsampled to 256 $Hz$, and a notch filter at 50 $Hz$ was applied to avoid the influence of power line interference.

Both datasets have the same acquisition paradigm. Subjects sat on a chair and a screen was in front of the subjects. EEG signals were acquired from the channels on the brain scalp. The channels used in the classification include $FC_z$, $C_3$, $C_z$, $C_4$, $CP_z$, $F_3$, $F_z$, $F_4$, $P_3$, $P_z$ and $P_4$. At the start of a trial, the screen displayed a cross. Two seconds later, a cue appeared on the screen indicating a motion of the upper limb which the subjects were then supposed to execute. In dataset I, the executed motions include \textit{elbow flexion}, \textit{elbow extension}, \textit{supination}, \textit{pronation}, \textit{hand open}, \textit{hand close} and the \textit{resting} state. Dataset II includes \textit{supination}, \textit{pronation}, \textit{hand open}, \textit{palmar grasp} and \textit{lateral grasp}. The number of trials for each motion were 60 and 72 in datasets I and II, respectively. The numbers of subjects in both datasets were 15 and 9, respectively.

Although both datasets have the same paradigm, the time windows of the EEG signals in the two datasets are different. In dataset I, the hand trajectory was simultaneously acquired along with the EEG signals. The movement onset of the executed motions can be located by the hand trajectory. The time window in which the EEG signals will be used for classification purposes lies between two seconds before the onset and one second after the onset. In dataset II, however, the hand trajectory was not recorded and there is no information about the movement. Therefore, here the time window for the classification corresponds to the two-second window after the cue indicating the start of the executed motions.

The movement onset is located with the movement trajectory in dataset I. The same localization process as in \cite{jia_improve_2022} was adopted. The localization of the movement is associated with the hand trajectory. The 1-order difference of the hand trajectory was filtered by a 1-order \textit{Savitzky-Golay} finite impulse response smoothing filter. The time window length in the smoothing filter was set to 31. The hand trajectory was then normalized by dividing by the maximal absolute value. In \textit{elbow flexion} and \textit{elbow extension}, the hand trajectory has a higher amplitude when the limb moves. The location where the normalized trajectory equals the threshold of 0.05 was the movement onset. Trials were manually removed if the movement onset could not be located because of noise contamination. In the \textit{resting} state, a fake movement onset was set to 0.5 $s$ after the cue appeared on the screen. Trials in the \textit{resting} state were rejected if the variances of normalized trajectory were less than 0.02. In the other four motions, the function $f(x)=a*exp(-(\frac{x-b}{c})^2)+d$ was used to fit the smoothed and normalized trajectory by tuning the parameters $a,b,c,d$. Trials were then rejected if the parameters of the tuned function met at least one of the following conditions: $a<0.05$, $c>100$ or $d>10$. The movement onset was set to the time point whose amplitude equalled 0.1. The average number of trials of each motion across subjects are given in Table \ref{tab:1}.

\begin{table}
        \caption{Number of Trials Across Subjects After Trial Rejection in Dataset I}
        \begin{tabular}[htbp]{c|c|c|c|c}
                \toprule
                Motion & \textit{elbow flexion} & \textit{elbow extension} & \textit{supination} & \textit{pronation} \\
                \midrule
                Number & 60 & 59 & 52 & 51 \\
                \midrule
                Motion & \textit{hand close} & \textit{hand open} & \textit{resting} & \\
                \midrule
                Number & 56 & 55 & 59 & \\
                \bottomrule
        \end{tabular}
        \label{tab:1}
\end{table}

\subsection{Binary FBTRCA}
\label{ssec:bFBTRCA}

The proposed multi-class FBTRCA (mFBTRCA) is developed based on the binary FBTRCA (bFBTRCA). To present the relationship these two methods, we first introduce the structure of bFBTRCA and then detail how mFBTRCA is developed based on bFBTRCA.

\begin{figure*}[htbp]
    \centering
    \includegraphics[width=0.75\textwidth]{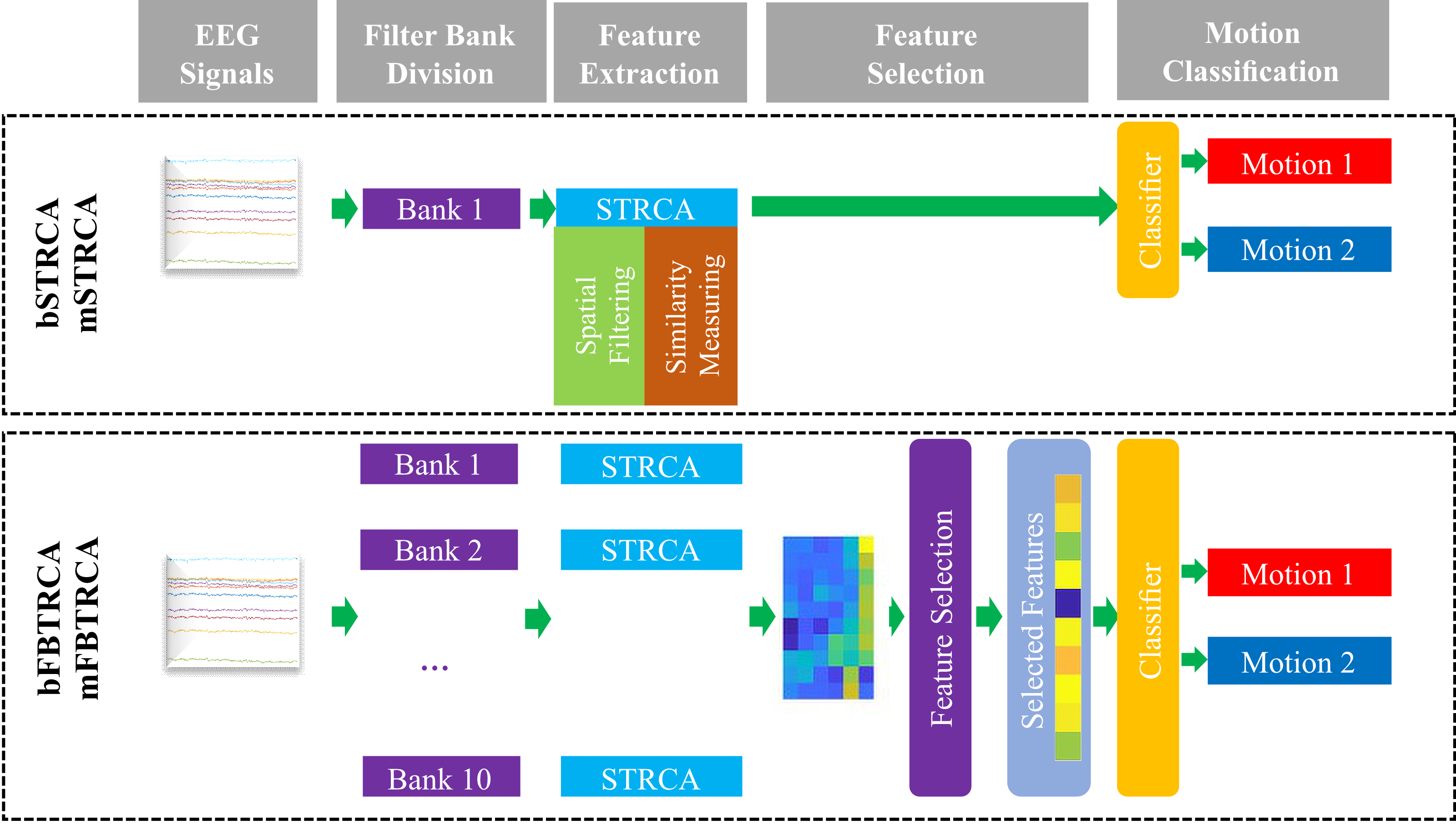}
    \caption{Relationship between the structure of bSTRCA and bFBTRCA. The STRCA in this figure is either the bSTRCA or the mSTRCA. Both bSTRCA and mSTRCA have two steps: spatial filtering and similarity measurement. The bFBTRCA or mFBTRCA is developed by applying bSTRCA/mSTRCA to multiple banks and enabling feature selection on the features from these banks.}
    \label{fig:2.1}
\end{figure*}

The bFBTRCA method is developed by incorporating filter bank selection into the bSTRCA method. The key idea of bFBTRCA is to find the best frequency band in which the bSTRCA method has the best classification performance. Instead of selecting the best frequency band, bFBTRCA selects the best features from features in all frequency bands. The bFBTRCA method first divides EEG signals into multiple filter banks in the low-frequency domain. In each filter bank, bSTRCA calculates the canonical correlation pattern and uses them as the features. In bFBTRCA, the features extracted using bSTRCA in all filter banks are sorted and selected through the minimum redundancy maximum relevance method. Finally, the support vector machine classifies the selected features. The relationship between bSTRCA and bFBTRCA is shown in Fig. \ref{fig:2.1}. The bFBTRCA method has three key points in the classification: (1) the spatial filtering with task-related component analysis, (2) the feature extraction from the canonical correlation pattern, and (3) filter bank selection.

\subsubsection{Spatial Filtering}
\label{sssec:btask}
Because EEG signals are multi-channel signals and the channels are isolated points on the brain scalp, EEG signals naturally have a bad spatial quality. Spatial filtering is commonly used to optimize the spatial quality when processing EEG signals. A spatial filter is used to find a matrix $\boldsymbol{W}\in\mathbb{R}^{C\times P}$, where $C$ is number of channels and $P\leq C$. By multiplying the given EEG signal $\boldsymbol{X}\in\mathbb{R}^{C\times T}$ with the matrix $\boldsymbol{W}$, the spatial-filtered signal $\boldsymbol{X}^T\boldsymbol{W}\in\mathbb{R}^{T\times P}$ is obtained. Here, $T$ is the number of sample points. Compared to the original EEG signal $\boldsymbol{X}$, the dimension of the spatial-filtered signals $\boldsymbol{X}^T\boldsymbol{W}$ is suppressed. The calculation methods of the spatial filter differ for different brain activities. For example, CSP is widely used in motor imagery, which aims to discern between the channels with the biggest and smallest variances \cite{zhang_temporally_2019}. In steady-state visual evoked potentials, the discriminative canonical pattern matching method is used to maximize the between-class difference and minimize the within-class difference \cite{xu_braincomputer_2018}. In the bSTRCA and bFBTRCA methods, the task-related component analysis is used as the spatial filter, thereby extracting the task-related components.

Task-related component analysis optimizes the signals based on covariances of the EEG signals. The training set of EEG signals is given as $\mathcal{X}_k=\{\boldsymbol{X}_1^k, \boldsymbol{X}_2^k,\dots,\boldsymbol{X}_{I_k}^k\}$, where $k$ is the index of classes; for instance, in binary classification, $k=1,2$. $I_k$ represents the number of trials of class $k$. $\boldsymbol{X}$ are multi-channel EEG signals of size $C\times T$. The task-related component analysis first computes the covariance of the self-trial and inter-trial of each class. The self-trial covariance is
\begin{equation}
        \boldsymbol{C}_{i}^k=\boldsymbol{X}_i^k(\boldsymbol{X}_i^k)^T,
        \label{eqn:1}
\end{equation}
while the inter-trial covariance is given by
\begin{equation}
        \boldsymbol{C}_{i,j}^k=\boldsymbol{X}_i^k(\boldsymbol{X}_j^k)^T+\boldsymbol{X}_j^k(\boldsymbol{X}_i^k)^T.
        \label{eqn:2}
\end{equation}
The spatial filter is the combination of eigenvectors, which is obtained by solving the following eigen equation:
\begin{equation}
        \max_{\boldsymbol{\omega}} J^k=\frac{\boldsymbol{\omega}^T \boldsymbol{S}^k\boldsymbol{\omega}}{\boldsymbol{\omega}^T \boldsymbol{Q}^k\boldsymbol{\omega}}.
        \label{eqn:3}
\end{equation}
$\boldsymbol{S}^k$ is the sum of inter-trial covariances of class $k$
\begin{equation}
        \boldsymbol{S}^k=\sum_{i,j=1,i<j}^{I_k}\boldsymbol{C}_{i,j}^k,
        \label{eqn:4}
\end{equation}
and $\boldsymbol{Q}^k$ is the sum of the self-trial covariances of class $k$
\begin{equation}
        \boldsymbol{Q}^k=\sum_{i=1}^{I_k}\boldsymbol{C}_{i}^k.
        \label{eqn:5}
\end{equation}
The eigen equation $\max_{\boldsymbol{\omega}} J^k$ can be solved with the generalized Schur decomposition as the generalized eigenvalue problem. The eigenvectors related to the maximal eigenvalues are denoted as $\boldsymbol{\omega}_k\in \mathbb{R}^{C\times P}$, where $P$ is the number of fetched eigenvectors. The spatial filter of task-related component analysis, $W$, is the concatenation of eigenvectors of two classes $\boldsymbol{W}=[\boldsymbol{\omega}_1,\boldsymbol{\omega}_2]\in \mathbb{R}^{C\times 2P}$. The optimized calculation step of the task-related component analysis can be found in \cite{tanaka_group_2020}.

\subsubsection{Similarity Measurement}
\label{sssec:bccp}
In MRCP signals, the grand average MRCP is the mean of EEG signals across trials, and is used to balance out the noise in each of the trials. It is denoted as:
\begin{equation}
        \hat{\boldsymbol{X}}^k=\sum_{i=1}^{I_k}\boldsymbol{X}_{i}^k/I_k.
        \label{eqn:6}
\end{equation}
When measuring the relationship between the grand average MRCP, $\hat{\boldsymbol{X}}^k$, and each of the trials $\boldsymbol{X}\in \mathbb{R}^{C\times T}$, bSTRCA and bFBTRCA both use the canonical correlation pattern to measure the similarity. The canonical correlation pattern includes three correlation coefficients:

(1) Correlation between $\boldsymbol{X}$ and $\hat{\boldsymbol{X}}^k$:
\begin{gather}
        \boldsymbol{X}_*=\boldsymbol{X};\boldsymbol{X}_k=\hat{\boldsymbol{X}}^k;
        \label{eqn:7}
        \\
        \rho_{1,k}=corr(\boldsymbol{X}_*^T\boldsymbol{W},\boldsymbol{X}_k^T\boldsymbol{W});
        \label{eqn:8}
\end{gather}

(2) Correlation between $\boldsymbol{X}$ and $\hat{\boldsymbol{X}}^k$ after canonical correlation analysis:
\begin{gather}
        \boldsymbol{X}_*=\boldsymbol{X};\boldsymbol{X}_k=\hat{\boldsymbol{X}}^k;
        \label{eqn:9}
        \\
        [\boldsymbol{A}_k,\boldsymbol{B}_k]=cca(\boldsymbol{X}_*^T\boldsymbol{W},\boldsymbol{X}_k^T\boldsymbol{W});
        \label{eqn:10}
        \\
        \rho_{2,k}=corr(\boldsymbol{X}_*^T\boldsymbol{W}\boldsymbol{B}_k,\boldsymbol{X}_k^T\boldsymbol{W}\boldsymbol{B}_k);
        \label{eqn:11}
\end{gather}

(3) Correlation between $\boldsymbol{X}-\hat{\boldsymbol{X}}^{k}$ and $\hat{\boldsymbol{X}}^{3-k}-\hat{\boldsymbol{X}^k}$ after canonical correlation analysis:
\begin{gather}
        \boldsymbol{X}_*=\boldsymbol{X}-\hat{\boldsymbol{X}}^k;\boldsymbol{X}_k=\hat{\boldsymbol{X}}^{3-k}-\hat{\boldsymbol{X}^k};
        \label{eqn:12}
        \\
        [\boldsymbol{A}_k,\boldsymbol{B}_k]=cca(\boldsymbol{X}_*^T\boldsymbol{W},\boldsymbol{X}_k^T\boldsymbol{W});
        \label{eqn:13}
        \\
        \rho_{3,k}=corr(\boldsymbol{X}_*^T\boldsymbol{W}\boldsymbol{A}_k,\boldsymbol{X}_k^T\boldsymbol{W}\boldsymbol{A}_k);
        \label{eqn:14}
\end{gather}

In the above equations, $corr$ corresponds to the two-dimensional Pearson correlation coefficient, and the function symbol $cca$ computes the canonical coefficients for the two input data matrices. Because the canonical correlation analysis is used in the feature extraction, the EEG signals must be z-normalized before spatial filtering in bSTRCA. The correlations between EEG signals $\boldsymbol{X}$ and the grand average MRCPs of two classes are calculated in the binary classification. The number of correlation features is six. 

\subsubsection{Filter Bank Selection}
\label{sssec:bfbselect}
In bFBTRCA, the filter bank selection consists of two steps: filter bank division and feature selection. After the z-normalization of the original EEG signals, these are divided into subbands in the low-frequency domain. The low cut-off frequencies of these subbands are fixed to 0.5 $Hz$, while their high cut-off frequencies are in the arithmetic sequence going from 1 $Hz$ to 10 $Hz$ with a 1 $Hz$ step. Therefore, ten filter banks are used in our work.

bSTRCA is used to extract features from each of the filter banks. This feature extraction includes spatial filtering and correlation coefficient extraction. The number of features is 6 in each subband, giving a total of 60 features from all subbands.

The adopted feature selection method is the minimum-redundancy maximum-relevance, which is used to select essential features from the total 60 features. Mutual information measures the mutual dependence between two variables, and it quantifies the information from one variable by observing the other variable. In the minimum-redundancy maximum-relevance, relevance is the mutual information between the label and the features, while redundancy is the mutual information between two features. The minimum-redundancy maximum-relevance method optimizes the sequence of features by minimizing the redundancy and maximizing the relevance. The selected features are then classified with the binary support vector machine classifier.

\subsection{Multi-class FBTRCA}
\label{ssec:mFBTRCA}
The bFBTRCA method is designed to classify two states of limb movement based on the differences between grand average MRCPs of two states. When adapting the bFBTRCA to solve the multi-class classification problem, the spatial filter's structure restricts the framework's extension. 

In the spatial filtering of binary classification, eigenvectors $\boldsymbol{\omega}_1\in \mathbb{R}^{C\times P}$ and $\boldsymbol{\omega}_2\in \mathbb{R}^{C\times P}$ of two classes are concatenated into the spatial filter $\boldsymbol{W}\in \mathbb{R}^{C\times 2P}$ used in bFBTRCA. In the $K$-class classification, the size of the spatial filter $\boldsymbol{W}$ is $\in \mathbb{R}^{C\times KP}$, where $K$ is the number of classes. In this case, the number of channels $KP$ after spatial filtering is greater than the number of channels $C$ of the original EEG signals. After having been filtered with $\boldsymbol{W}\in \mathbb{R}^{C\times KP}$, the EEG signals are not full-rank, and therefore contain more redundant information than the original EEG signals before spatial filtering. The framework of bFBTRCA is optimized to fit with the multi-class classification.

\begin{figure*}[htbp]
    \centering
    \includegraphics[width=0.75\textwidth]{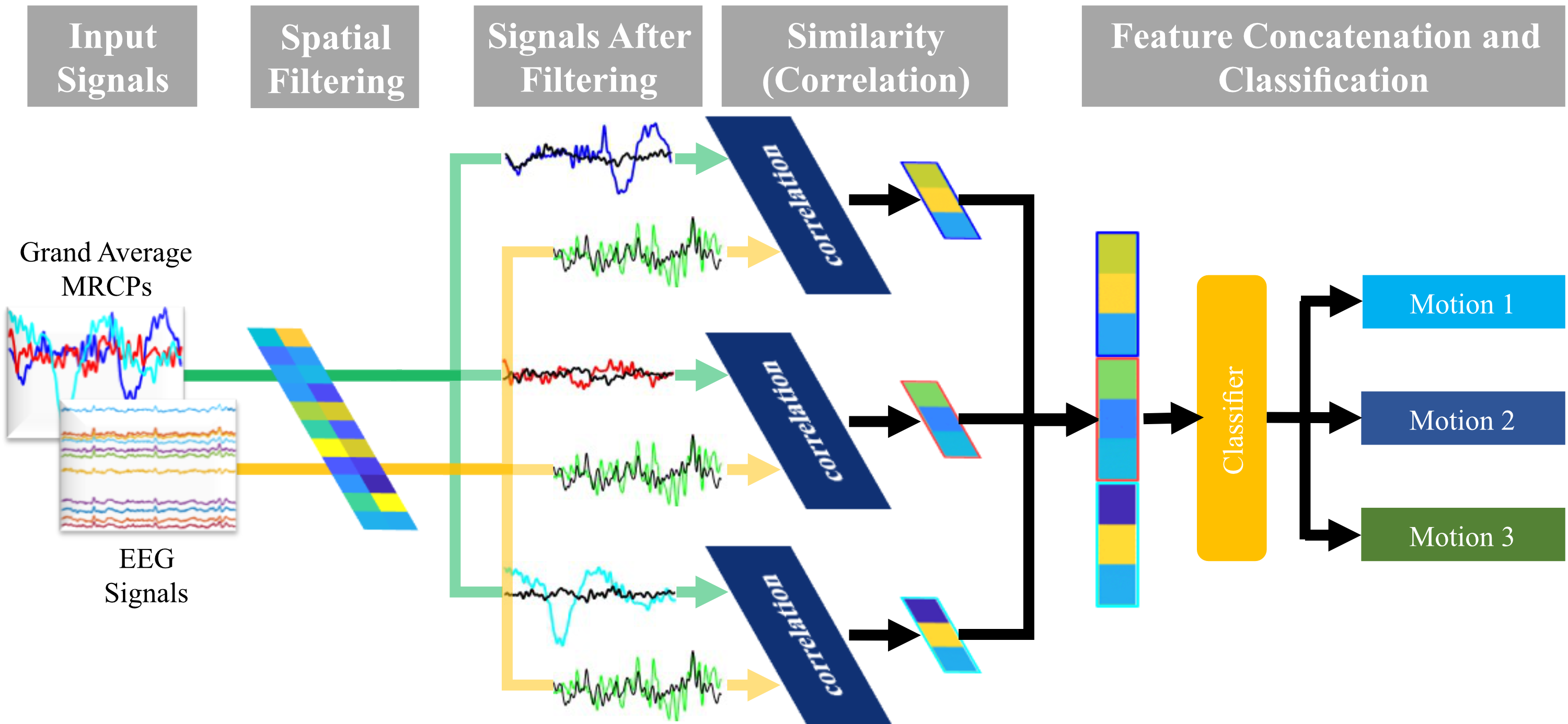}
    \caption{Structure of the mSTRCA method for the three-class classification problem. This structure can be extended to a classification model for $K$ classes, where $K \in \mathbb{Z}^{+}$. The optimization of the bSTRCA includes two main points. The first is the optimization of the spatial filter, which avoids the dimensional increase of EEG signals in multi-class classification. The second is the input of the correlation. The same components are removed from the signals after spatial filtering when measuring the similarity between the grand average MRCPs and the EEG signals, as given in Equation \ref{eqn:16}. The proposed mFBTRCA method incorporates the filter bank selection into mSTRCA. This procedure is given in Fig. \ref{fig:2.1}.}
    \label{fig:2.2}
\end{figure*}

This optimization includes two points: the spatial filter and the similarity measurement. After the optimization, the bSTRCA method can be used in the multi-class classification, which is the multi-class standard task-related component analysis (mSTRCA) method. The mFBTRCA method is developed by applying the filter bank selection to the mSTRCA method. The structure of the mSTRCA is shown in Fig. \ref{fig:2.2}, using three-class classification as an example.

\subsubsection{Spatial Filtering}
\label{sssec:mtask}
The bSTRCA is a method developed based on the grand average MRCP. Before optimizing the frame of bSTRCA, it is necessary to clarify the relation between bSTRCA and the grand average MRCP. The grand average MRCP is the mean of EEG trials in the same class. 
There are three kinds of inputs involved in the calculation of the correlation coefficients in bSTRCA:

(1) each of the EEG trials before averaging

(2) the grand average MRCP of one class

(3) the grand average MRCP of the other class.

The binary classification is based on the differences between the two grand average MRCPs. The features in bSTRCA use the similarities between each EEG trial as well as the two grand average MRCPs with correlation coefficients. The labels of the EEG trials can be predicted by their similarity. However, the noise in EEG signals are not eliminated by taking the mean of all trials. bSTRCA uses the task-related component analysis as a spatial filter to reject the task-unrelated components such as noise from the original EEG signals. Therefore, the spatial filter plays a main role in rejecting noise here, and is not related to discriminating the classes of EEG signals. 

In bSTRCA, the eigenvectors of two classes are obtained by solving the eigen equation in Equation \ref{eqn:3} and are then concatenated into the used spatial filter. However, it is not necessary to label the eigenvectors in the spatial filter, because the filter is used for noise rejection and task-related component extraction. Since the spatial filter is responding to the noise rejection and is not related to the classification, the spatial filter used in bSTRCA is modified such that it only removes the information about classes in the spatial filter. 

The summed-up inter-trial covariance $\boldsymbol{S}^k$ and the summed-up self-trial covariance $\boldsymbol{Q}^k$ are obtained through Equations \ref{eqn:4} and \ref{eqn:5}. The spatial filter for the multi-class classification is found using the eigen equation

\begin{equation}
        \max_{\boldsymbol{\omega}} J=\frac{\boldsymbol{\omega}^T \boldsymbol{S} \boldsymbol{\omega}}{\boldsymbol{\omega}^T \boldsymbol{Q} \boldsymbol{\omega}},
        \label{eqn:15}
\end{equation}
where $\boldsymbol{S}=\sum_{k=1}^K \boldsymbol{S}^k$ and $\boldsymbol{Q}=\sum_{k=1}^K \boldsymbol{Q}^k$. $K$ is the number of classes in the multi-class classification.

\subsubsection{Similarity Measurement}
\label{sssec:mccp}
In binary classification, the performance is determined by the differences between the grand average MRCP of two motions. When two grand average MRCPs have large differences, this indicates that the classification accuracy of the two motions is higher than that with minor differences.
To reduce the similarity between two grand average MRCPs, a possible approach is to remove the mean of the two from both grand average MRCPs. The differences between to grand average MRCPs are then maximized. In the multi-class task, the mean grand average MRCPs of $K$ motions are removed from the grand average MRCP $\hat{\boldsymbol{X}}^k$ and the input EEG signals $\boldsymbol{X}$ in Equation \ref{eqn:7}, \ref{eqn:9} and \ref{eqn:12}
\begin{equation}
        \boldsymbol{X}\to\boldsymbol{X}-\frac{1}{K}\sum_{k=1}^{K}\hat{\boldsymbol{X}}^k;
        \hat{\boldsymbol{X}}^k\to\hat{\boldsymbol{X}}^k-\frac{1}{K}\sum_{k=1}^{K}\hat{\boldsymbol{X}}^k.
        \label{eqn:16}
\end{equation}
The canonical correlation pattern consists of three correlation coefficients for each class in multi-class classification. The first two correlation coefficients are the same as the ones given in Equations \ref{eqn:8} and \ref{eqn:11}. The third one is given by Equation \ref{eqn:14}. However, to fit with the needs of multi-class classification, Equation \ref{eqn:12} is replaced with
\begin{equation}
        \boldsymbol{X}_*=\boldsymbol{X}-\hat{\boldsymbol{X}}^k;\boldsymbol{X}_k=\frac{1}{K-1}\sum_{kk=1,kk\neq k}^{K}\hat{\boldsymbol{X}}^{kk}-\hat{\boldsymbol{X}^k}.
        \label{eqn:17}
\end{equation}

In Equation \ref{eqn:12}, $\boldsymbol{X}_k$ is the distance between $\hat{\boldsymbol{X}}^1$ and $\hat{\boldsymbol{X}}^2$. The distances between $\boldsymbol{X}$ and the grand average MRCPs, namely $\boldsymbol{X}-\hat{\boldsymbol{X}}^1$ and $\boldsymbol{X}-\hat{\boldsymbol{X}}^2$, are normalized by the correlation with $\hat{\boldsymbol{X}^1}$ and $\hat{\boldsymbol{X}^2}$. In the multi-class classification, $\hat{\boldsymbol{X}}^k$ is given in Equation \ref{eqn:17} to normalize the distance $\boldsymbol{X}_*$. The number of coefficients for each class is three, and the similarity is measured with $3K$ coefficients in total.

\subsubsection{Filter Bank Selection}
\label{sssec:mfbselect}
In the filter bank selection of mFBTRCA, the same setting is used as the one presented in Section \ref{sssec:bfbselect}. EEG signals are divided into ten filter banks, and the minimum-redundancy maximal-relevance method is used to optimize the sequence of features and select the best features for classification. The selected features are classified using the multi-class support vector machine method.

\subsection{Comparison Methods}
\label{ssec:benchmark}
\subsubsection{State-of-the-Art methods}
\label{sssec:sota}
The following is a brief introduction to the compared state-of-the-art models on the multi-class classification of limb movements. All the given methods were implemented on the same datasets used in this paper.
\paragraph{mCSP+LDA\cite{grosse-wentrup_multiclass_2008}} 
The multi-class CSP (mCSP) is an extension of the binary CSP used in motor imagery. In mCSP, EEG signals are optimized with joint approximate diagonalisation, and independent components are chosen by maximizing mutual information of the independent components and class labels. The pre-processed results are fed into the linear discriminant analysis classifier for class label prediction. 

\paragraph{SpoC+Ridge\cite{dahne_spoc_2014}}
The source power comodulation (SPoC) method can be seen as an extension of CSP that allows to extract spatial filters and patterns by using a continuous target variable in the decomposition process. The use of the target variable gives preference to components whose power correlates with the target variable. Features extracted by SPoC are classified with the ridge regression classifier.

\paragraph{MDM\cite{barachant_multiclass_2012}}
The minimum distance to mean (MDM) uses the Riemannian distance and Riemannian mean to classify the signals. It first converts the multi-channel signals into self-covariances and casts these to points on the Riemannian manifold. It then uses the Riemannian mean to obtain the center point of points belonging to a class, and the Riemannian distance to measure the distance between the center points and the unlabelled points. The Riemannian distance is used to predict the class of the unlabelled trials.

\paragraph{TSLDA\cite{barachant_multiclass_2012}}
The first step of the tangent space linear discriminant analysis (TSLDA) method is also to convert EEG signals into covariances. It then maps the covariance matrices onto the Riemannian tangent space. The covariance matrices are vectorized and treated as Euclidean objects. Finally, the linear discriminant analysis classifier is applied to the dimension-reduced features.

\paragraph{SCNN and DCNN\cite{schirrmeister_deep_2017}} CNN has revolutionized computer vision through learning from raw data. Its use has been studied on decoding executed or imagined tasks from raw EEG by shallow CNN (SCNN) without handcrafted features. Compared to SCNN, deep CNN (DCNN) has a deep network architecture and a better fitting capability.

\paragraph{WaveNet\cite{thuwajit_eegwavenet_2022}}
The EEG WaveNet is a multi-scale CNN that works for the detection of epileptic seizures. This network consists of trainable depth-wise convolutions and spatial-temporal convolutions. Because seizures are related to movement, the performance of this network is used as a comparison method in this work.

\paragraph{HopeFullNet\cite{mattioli_1d_2021}} The HopeFullNet is a one-dimensional CNN used in the classiﬁcation task of motor imagery. This network has shown its state-of-the-art performance in classifying four imagined movements and the resting state. The imagined movements included the opening and closing of (1) the left fist, (2) the right fist, (3) both fists, and (4) both feet.

\subsubsection{Baseline Methods}
\label{sssec:baseline}
As EEG signals are multi-channel time series, it is unavoidable to discuss their temporal characteristics. RNN is a universal solution to the feature extraction of time series. Along with the state-of-the-art methods presented above, our model is also compared with models that combine the RNN and CNN layers. In previous EEG signal analyses, CNN was used to extract features from EEG data; the extracted features were then processed by the RNN layers to extract the temporal features, and and finally the fully connected layer was used as the classifier \cite{xu_one-dimensional_2020,shi_hybrid_2019,iyer_cnn_2022}. The baseline method, convolutional-RNN-convolutional neural network (C-R-CNN), follows the steps in the previous analysis and includes three modules: 

\noindent Module 1: CNN layers used to optimize spatial characteristics

\noindent Module 2: RNN layers used to capture temporal characteristics

\noindent Module 3: CNN layers followed by a fully connected layer used to cover overall features and classify them. 

The model structure of the baseline method is given in Table \ref{tab:2}, where the batch size is $B$ and the number of classes is $K$. The sequence of the output size is $[batch, channel, high, width]$.

\begin{table}[htbp]
    \centering
    \caption{The model structure of the baseline method}
    \begin{tabular}{c|c|c}
        \toprule
        Module & Layer & Output Size \\
        \midrule
        Input & & [$B$, 1, $C$, $T$]\\
        \midrule
        \multirow{7}{*}{Module 1} & ZeroPad2d & [$B$, 1, $C$, $T$+31]\\
        & Conv2d & [$B$, 8, $C$, $T$]\\
        & BatchNorm2d & [$B$, 8, $C$, $T$]\\
        & LeakyReLU & [$B$, 8, $C$, $T$]\\
        & Conv2d & [$B$, 16, 1, $T$]\\
        & BatchNorm2d & [$B$, 16, 1, $T$]\\
        & LeakyReLU & [$B$, 16, 1, $T$]\\
        \midrule
        \multirow{2}{*}{Module 2} & Permutation  & [$B$, $T$, 16]\\
        & RNN & [$B$, 1, $T$, 32]\\
        \midrule
        \multirow{5}{*}{Module 3} & Conv2d & [$B$, 64, $T$//256, 1] \\
        & BatchNorm2d & [$B$, 64, $T$//256, 1] \\
        & Flatten & [$B$, 64*$T$//256]\\
        & Linear & [$B$, 32] \\
        & LeakyReLU & [$B$, 32]\\
        & Linear & [$B$, 32]\\
        \bottomrule
    \end{tabular}
    \label{tab:2}
\end{table}

The RNN layer in Table \ref{tab:2} can also be replaced by either the gate recurrent unit (GRU) layer or the long short-term memory (LSTM) layer to achieve an improved performance. Therefore, we have three baseline methods: (1) C-R-CNN, (2) the convolutional-GRU-convolutional neural network (C-G-CNN) and (3) convolutional-LSTM neural network (C-L-CNN). The models of C-G-CNN and C-L-CNN correspond to the model that replaces the RNN layer of C-R-CNN with the GRU and LSTM layers, respectively. The three baseline methods share the same network architecture.

GNN is the other approach to optimizing the spatial characteristics of EEG signals. The channels of the signals are regarded as nodes of the graph. GNN optimizes the spatial quality of EEG signals by learning about the relation between these nodes with graph knowledge. The latent correlation layer in StemGNN is used to optimize the EEG signals in the C-G-CNN model \cite{stemgnn}. The attention mechanism in the latent correlation layer is used to learn the latent correlations between the multi-channel time series. The learned attention graph is passed to the 4-order Chebyshev polynomial; thus, the output graph is of size [$4$, $C$, $C$]. The input EEG signals of size [$B$, 1, $C$, $T$] are weighted with the learned graph. Therefore, the size of the latent correlation layer is [$B$, 4, $C$, $T$]. The EEG signals optimized with the latent correlation layer are then passed to the C-G-CNN model, with the $channel$ changed to 4. This baseline model is referred to as Graph C-G-CNN (GC-G-CNN) in the following sections.

\subsubsection{Parameter Settings}
The performance of the proposed and the compared methods are evaluated by 10-fold cross-validation. The compared methods contain various neural networks, including \textit{SCNN}\cite{schirrmeister_deep_2017}, \textit{DCNN}\cite{schirrmeister_deep_2017}, \textit{Wavenet}\cite{thuwajit_eegwavenet_2022}, \textit{HopeFullNet}\cite{mattioli_1d_2021} and the baseline methods. The compared neural networks have the same hyper-parameters, including batch size (50), learning rate (0.001) and training epochs (50). The loss function used is the cross-entropy, and the optimizer is the Adam algorithm. Both datasets are split based on the 10-fold cross-validation. The performance of all these methods is evaluated by the classification accuracy averaged from the 10 folds.

\section{Result}
\label{sec:result}
In this work, the mFBTRCA method is proposed to solve the multi-class classification problem of upper limb movements. Two datasets are used to evaluate and compare the proposed methods' performance against state-of-the-art and baseline methods. The results analysis consists of three parts: (1) the performance comparison between bFBTRCA and mFBTRCA in the binary classification task, (2) the evaluation of a three-class classification including two limb motions and the \textit{resting} state, and (3) the multi-class classification performance evaluation. The first and second parts are evaluated and analyzed with EEG signals in dataset I. In the third part, both datasets are used to analyze the relationship between the classification accuracy and the grand average MRCP of each motion.

\subsection{Structure Comparison}
\label{ssec:equalperf}
\begin{figure}[htbp]
    \centering
    \includegraphics[width=0.475\textwidth]{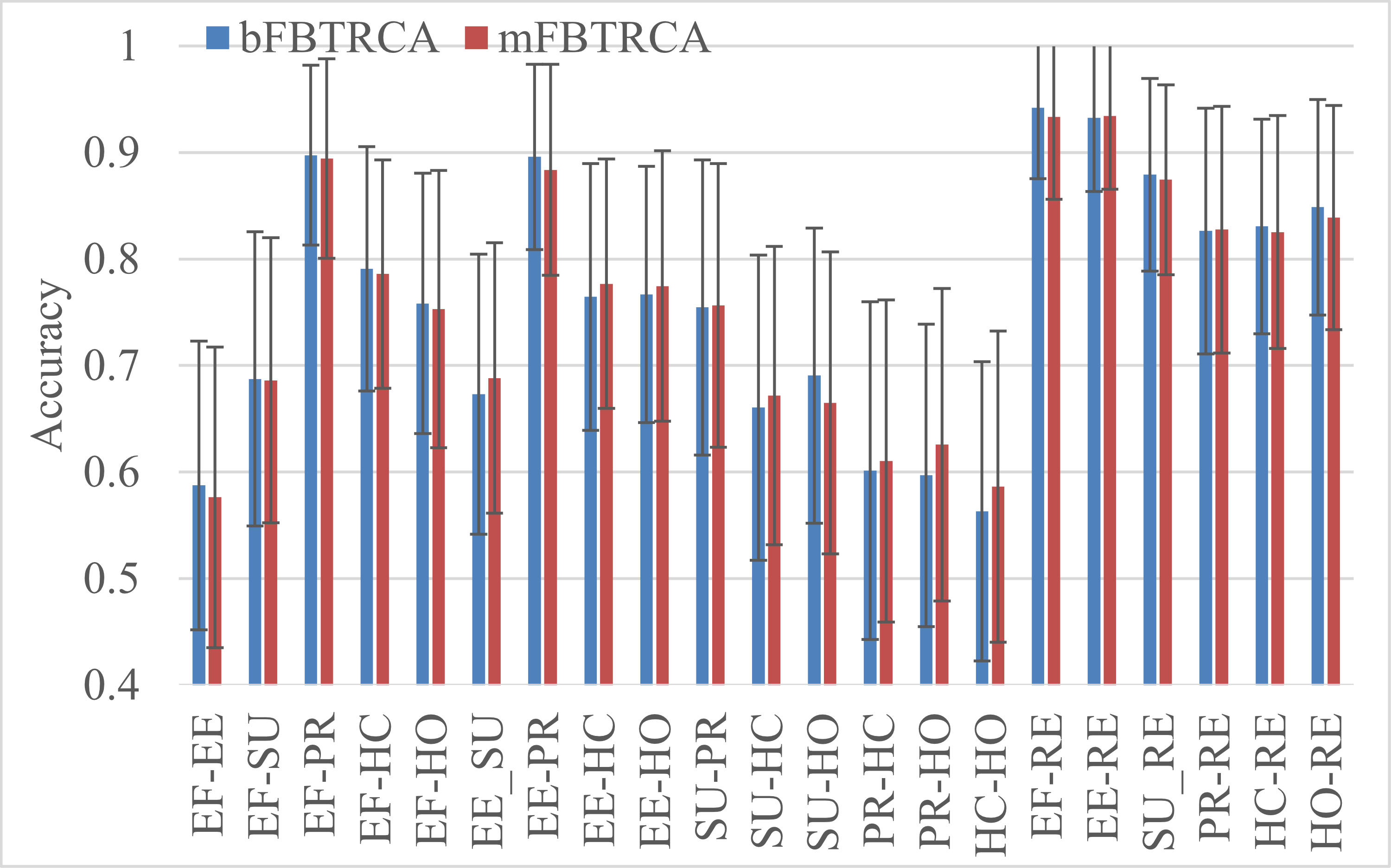}
    \caption{Comparison of the classification accuracies between the bFBTRCA and mFBTRCA methods in the binary classification task. The abbreviations on the x-axis refer to the motion names; for instance, 'EE' is the abbreviation for \textit{elbow extension}. The evaluation is based on the 21 motion pairs in dataset I. Accuracies are averaged across ten folds of 15 subjects. The mFBTRCA method performs similarly to bFBTRCA in binary classification.}
    \label{fig:3.1}
\end{figure}

This work optimizes the spatial filter and similarity measurement of bFBTRCA such that the resulting mFBTRCA method can be used in multi-class classification. Before applying mFBTRCA to multi-class classification tasks, it is necessary to compare the performances of bFBTRCA and mFBTRCA in the binary classification task. Therefore, the bFBTRCA and mFBTRCA methods are applied to classify motion pairs in dataset I. Fig. \ref{fig:3.1} gives the classification accuracies summarized from 10 folds of 15 subjects and 21 motion pairs. In the spatial filter of both bFBTRCA and mFBTRCA, the number of selected eigenvectors $P$ is 2 \cite{jia_improve_2022}. It can be seen that both methods achieve similar classification accuracies in the binary classification task.

\subsection{Three-class Comparison}
\label{ssec:threecls}
\begin{figure*}[htbp]
    \centering
    \includegraphics[width=\textwidth]{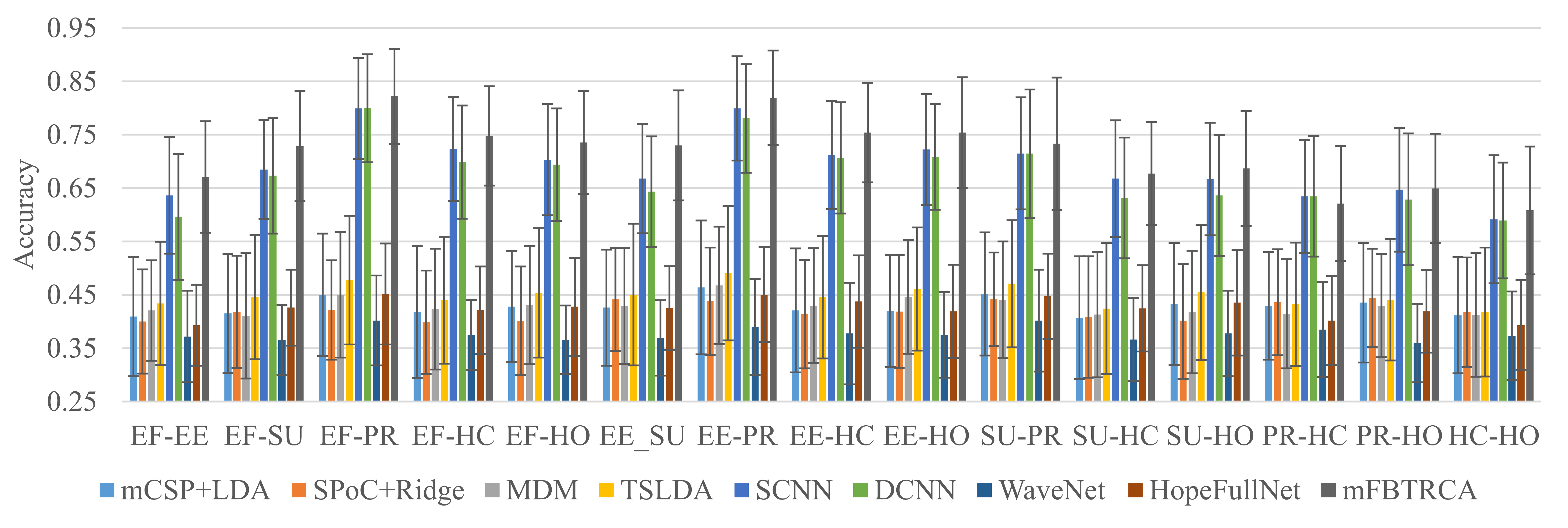}
    \caption{Three-class classification performance comparison between the proposed mFBTRCA method and the state-of-the-art methods. The three classes in this figure are the motion pairs of limb movements and the $resting$ state. The labels on the x-axis represent the name of the motion pairs, and the name for the the $resting$ state, 'RE', is ignored. The mFBTRCA method has a comparable performance to the SCNN and DCNN methods.}
    \label{fig:3.2}
\end{figure*}

This three-class comparison is carried out based on the classification accuracy between motion pairs of movement states and the $resting$ state. In Fig. \ref{fig:3.2}, the classification performances of the proposed mFBTRCA method and the state-of-the-art methods are given. As can be observed, the SCNN and DCNN methods have a comparable performance to mFBTRCA. However, the process going from EEG signals to the classification features in SCNN and DCNN is ambiguous due to the interpretability of neural networks; we know that the performance of the deep neural network is good, but do not know how the network utilizes the information in the EEG signals. For the proposed mFBTRCA method, on the other hand, it is clear how the MRCP signals are transformed into features. The method removes noise via the spatial filter and measures the similarity via the correlation coefficients as features.

\subsection{Multi-class Comparison}
\label{ssec:sevencls}

\begin{figure*}[htbp]
    \centering
    \subfigure[Dataset I with Movement Onset]{
    \includegraphics[width=0.25\textwidth]{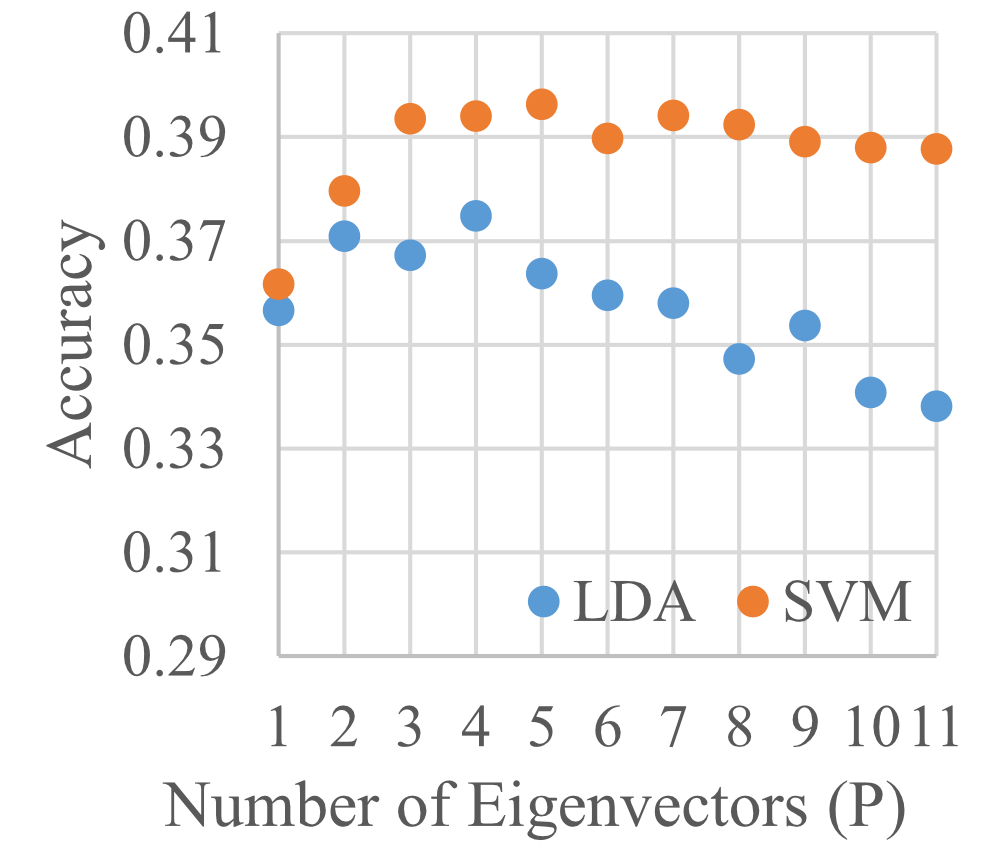}
    \label{fig:3.3a}
    }
    \subfigure[Dataset II without Movement Onset]{
    \includegraphics[width=0.25\textwidth]{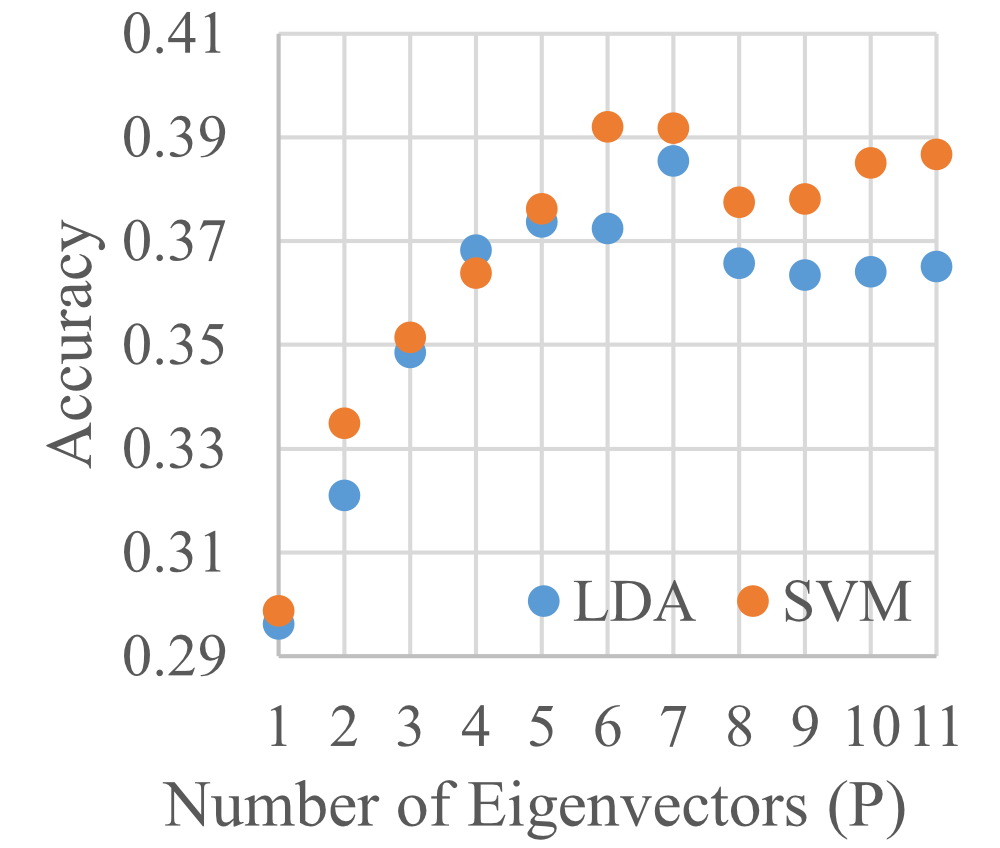}
    \label{fig:3.3b}
    }
    \subfigure[Dataset I without Movement Onset]{
    \includegraphics[width=0.25\textwidth]{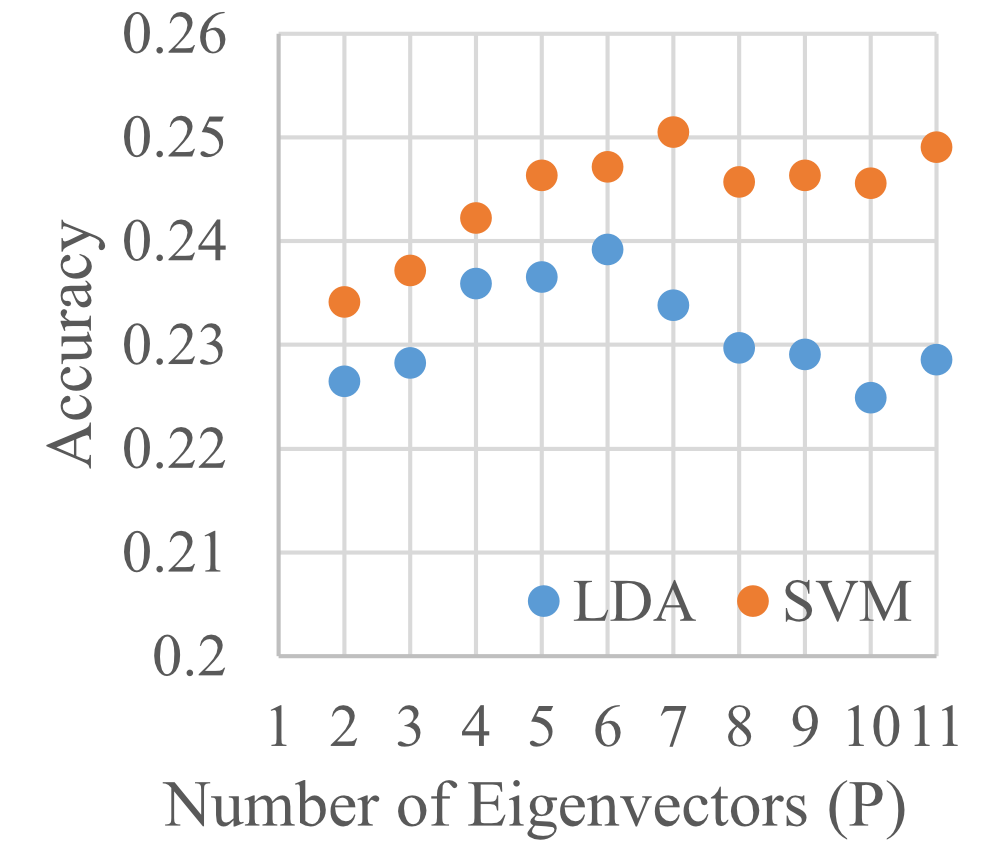}
    \label{fig:3.3c}
    }
    \caption{The mean of the classification accuracy of mSTRCA in datasets I and II. Dataset I includes the movement trajectory such that the movement onset can be located, but the same cannot be said for dataset II. The optimal number of eigenvectors in Fig. \ref{fig:3.3a} and Fig. \ref{fig:3.3b} is different. To explore the influence of the movement onset on $P$, the mSTRCA method is also applied to dataset I without locating the movement onset. As shown here, in both datasets I and II, when the movement onset is not located and aligned in the data pre-processing, the best $P$ is 6 or 7, whereas when the movement onset is located, the optimal $P$ value is 2 or 3.}
    \label{fig:3.3}
\end{figure*}

In the results analysis of the multi-class comparison, the performance of the proposed mFBTRCA method is compared to the state-of-the-art methods and the baseline methods with EEG signals from both datasets.

\paragraph{Number of Eigenvectors} The number $P$ of fetched eigenvectors in Equation \ref{eqn:15} may influence the classification performance. In both the binary and the three-class classifications, we followed the parameter settings used in our previous work \cite{jia_improve_2022} and set $P$=2. In the multi-class classification, we first compare the classification performance of mSTRCA with $P$ ranging from 1 to 11 (the number of channels). In Fig. \ref{fig:3.3a} and \ref{fig:3.3b}, the results of mSTRCA classified by the linear discriminant analysis (LDA) and support vector machine (SVM) models are given.

We assume that the supposed number of fetched eigenvectors $P$ is related to whether the movement onset is located and aligned in the dataset. In dataset I, the movement onset is located by the movement trajectory and EEG signals are fetched within two seconds before the movement onset and one second after that. The movement onsets of all the trials are aligned in dataset I. However, in dataset II, the movement onsets are not aligned because of the lack of movement trajectory. To validate this assumption, the mSTRCA method is applied to dataset I without locating the movement onset. The EEG signals are located within one second before the visual cue and two seconds after that. Subjects are supposed to execute the corresponding movement as the cue appears. The classification performance of mSTRCA in the non-aligned dataset I is given in Fig. \ref{fig:3.3c}. In this figure, the accuracy of mSTRCA also begins to decrease as $P$ becomes greater than 6. This phenomenon supports the assumption that $P$ is related to whether the movement onset is located or not. Based on the results presented in Fig. \ref{fig:3.3}, $P$ is set to 2 or 3 when the movement onset is located and aligned in the data pre-processing, and it is set to 6 or 7 when the movement onset cannot be located. As a result, $P$ is set to 3 and 6 for datasets I and II, respectively.

\paragraph{Overall Comparison} The overall performances of the proposed mFBTRCA method and the compared models are summarized in Table \ref{tab:3}, where the accuracies are averaged across all subjects and folds of each dataset. All the methods were applied to the same datasets (I and II). In dataset I, mFBTRCA improves on the classification accuracy of SCNN by 4.19\%. Furthermore, in dataset II, the proposed method improves on the classification accuracy of GC-G-CNN by 8.73\%.
\begin{table}[htbp]
    \centering
    \caption{Comparison with the state-of-the-art and baseline methods. All the methods are applied on the same datasets (I and II).}
    \begin{tabular}{c c|c|c}
         \toprule
         \multicolumn{2}{c|}{\multirow{2}{*}{Method}} & \multicolumn{2}{c}{Accuracy (Mean$\pm$Std)} \\
         \cmidrule{3-4}
         & & Dataset I (7 Classes) & Dataset II (5 Classes)\\
         \midrule
         mCSP+LDA &\cite{grosse-wentrup_multiclass_2008}& 0.1909$\pm$0.0584 & 0.2667$\pm$0.0797\\
         SPoC+Ridge &\cite{dahne_spoc_2014}& 0.1888$\pm$0.0492 & 0.2387$\pm$0.0562\\
         MDM &\cite{barachant_multiclass_2012}& 0.1947$\pm$0.0589 & 0.2768$\pm$0.0756\\
         TSLDA &\cite{barachant_multiclass_2012}& 0.2110$\pm$0.0605 & 0.2892$\pm$0.0686\\
         SCNN &\cite{schirrmeister_deep_2017}& 0.3603$\pm$0.0640 & 0.3083$\pm$0.0640\\
         DCNN &\cite{schirrmeister_deep_2017}& 0.3456$\pm$0.0671 & 0.2975$\pm$0.0748\\
         WaveNet &\cite{thuwajit_eegwavenet_2022}& 0.1597$\pm$0.0377 & 0.2327$\pm$0.0561\\
         HopeFullNet &\cite{mattioli_1d_2021}& 0.2025$\pm$0.0501 & 0.2292$\pm$0.0534\\
         \midrule
         \multicolumn{2}{c|}{C-R-CNN} & 0.3242$\pm$0.0687 & 0.2971$\pm$0.0705 \\
         \multicolumn{2}{c|}{C-L-CNN} & 0.3352$\pm$0.0672 & 0.3121$\pm$0.0748 \\
         \multicolumn{2}{c|}{C-G-CNN} & 0.3350$\pm$0.0755 & 0.3083$\pm$0.0694 \\
         \multicolumn{2}{c|}{GC-G-CNN} & 0.3069$\pm$0.0717 & 0.3159$\pm$0.0736 \\
         \midrule
         \multicolumn{2}{c|}{mFBTRCA} & $\boldsymbol{0.4022}\pm\boldsymbol{0.0709}$ & $\boldsymbol{0.4032}\pm\boldsymbol{0.0739}$\\
         \bottomrule
    \end{tabular}
    \label{tab:3}
\end{table}

\begin{figure*}[htbp]
        \centering
        \subfigure[Dataset I (7 classes)]{
        \includegraphics[width=0.475\textwidth]{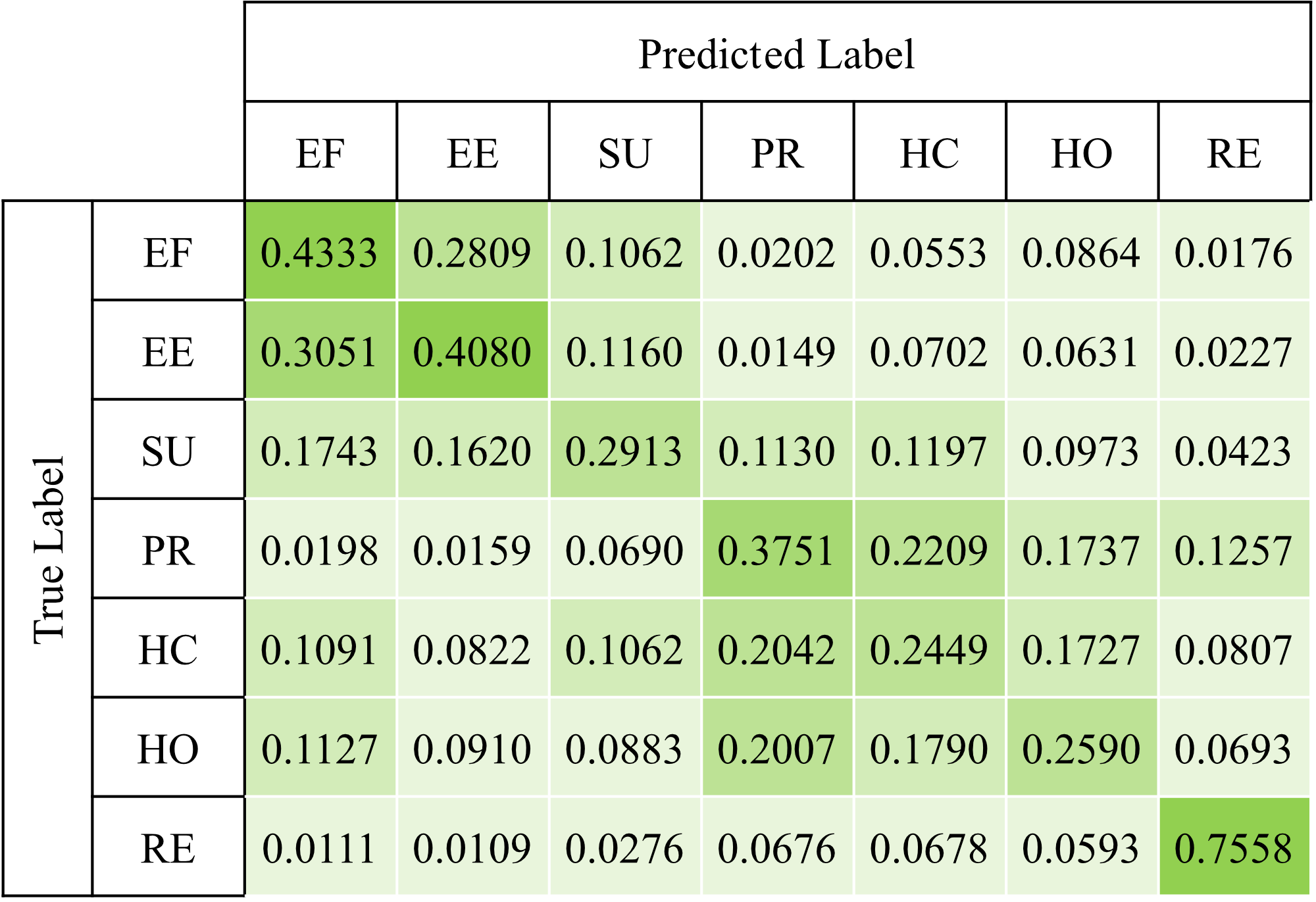}
        \label{fig:3.4a}
        }
        \subfigure[Dataset II (5 classes)]{
        \includegraphics[width=0.475\textwidth]{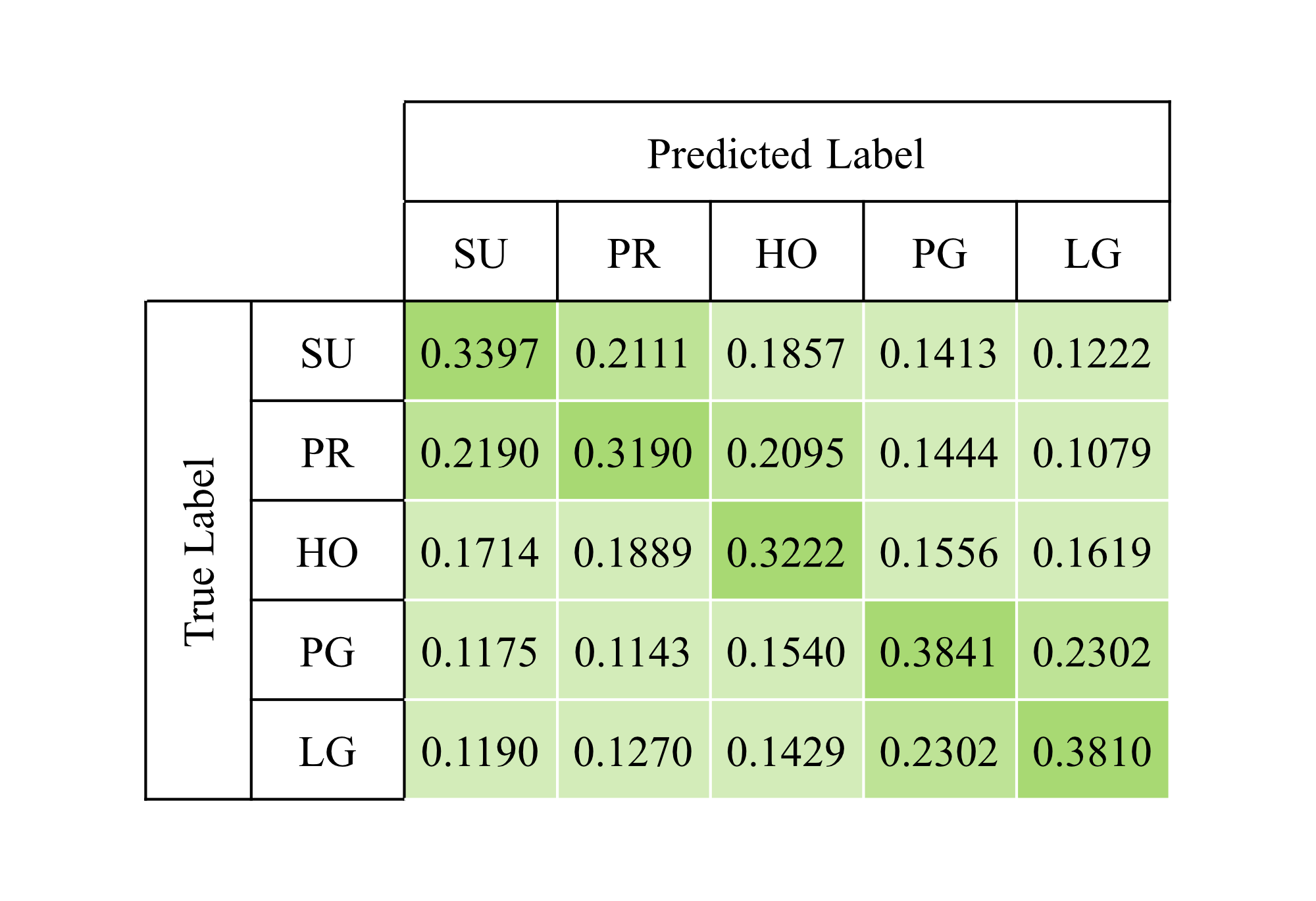}
        \label{fig:3.4b}
        }
        \caption{Confusion matrices in the multi-class classification task classified by mFBTRCA. The values given here are the confusion matrices having been normalized by dividing by the summed-up values in each row. Statistics are averaged from all subjects and folds in each dataset. In the confusion matrix of dataset I, it can be observed that the \textit{elbow flexion} and \textit{elbow extension} motion pair cannot be clearly discriminated; however, the \textit{resting} state is generally well separated from other states. The conclusion is consistent with the classification results given in Fig. \ref{fig:3.1} and Fig. \ref{fig:3.2}. In dataset II, the \textit{supination} and \textit{pronation} motion pair along with the \textit{palmar grasp} and \textit{lateral grasp} motion pair are also not well-separated.}
        \label{fig:3.4}
\end{figure*}

\begin{figure*}[htbp]
        \centering
        \subfigure[Dataset I (7 classes)]{
        \includegraphics[width=0.475\textwidth]{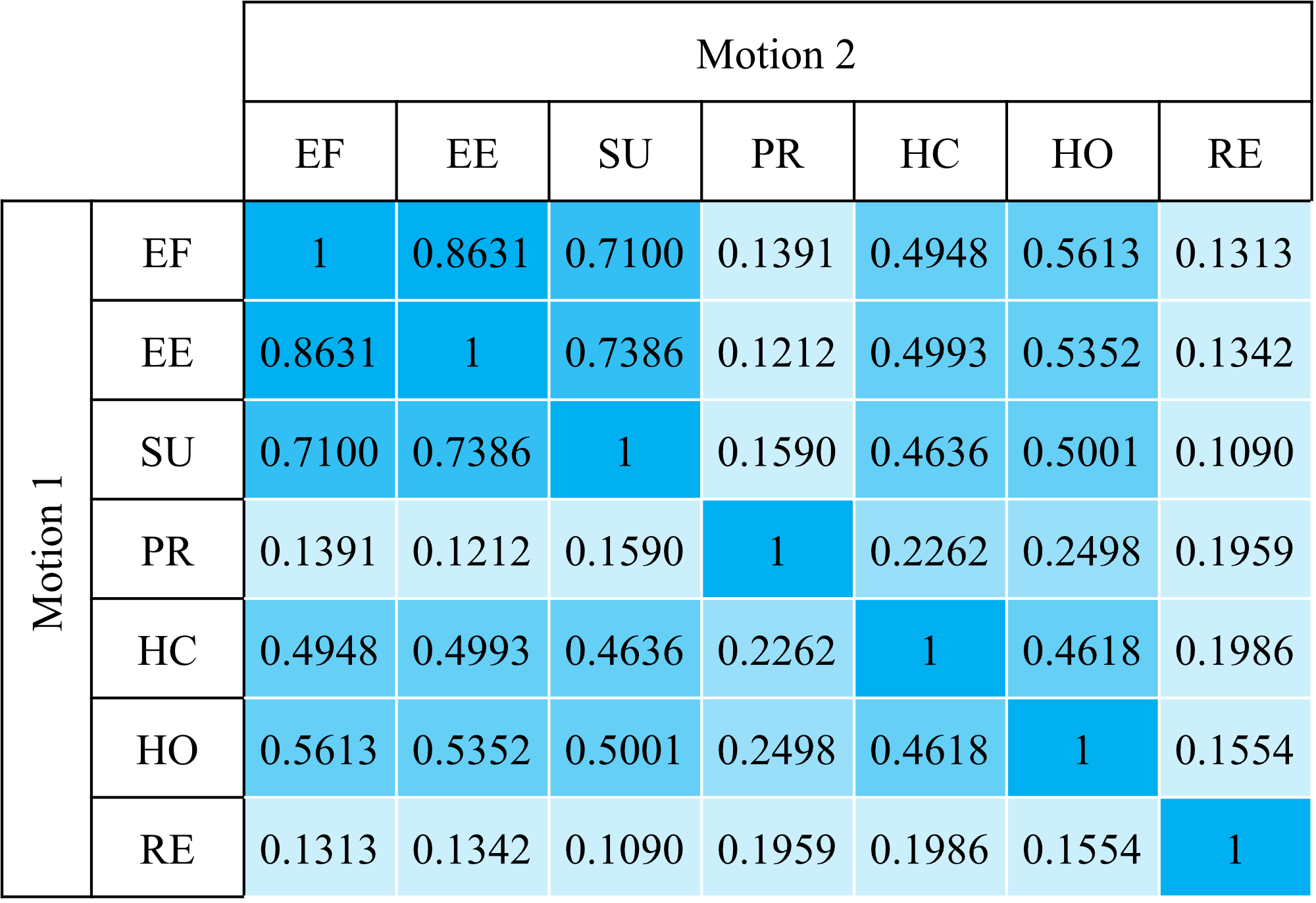}
        \label{fig:3.5a}
        }
        \subfigure[Dataset II (5 classes)]{
        \includegraphics[width=0.475\textwidth]{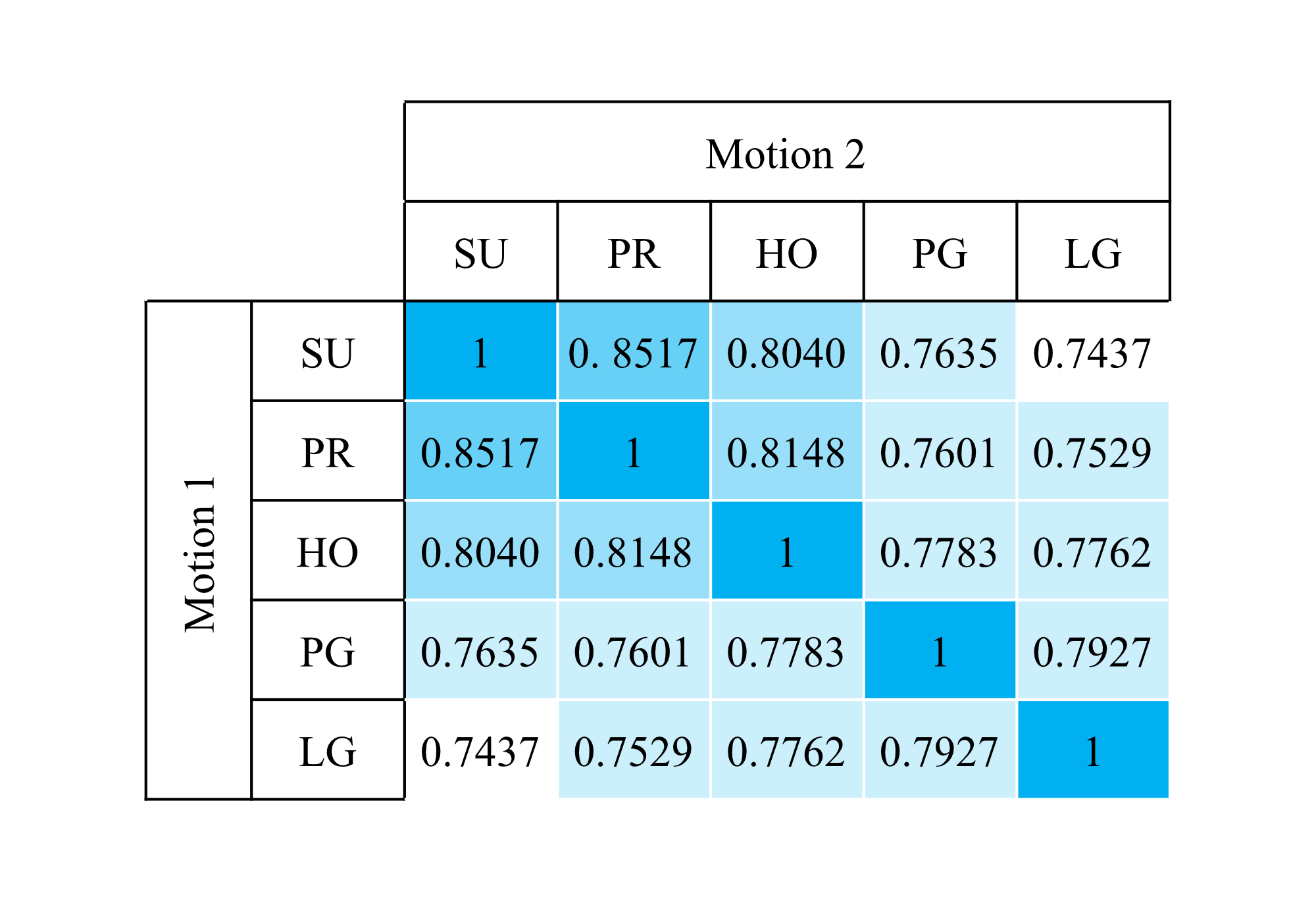}
        \label{fig:3.5b}
        }
        \caption{Two-dimensional Pearson correlation between the grand average MRCPs of two motions in the multi-class classification task, classified using mFBTRCA. Statistics are averaged from all subjects and folds in each dataset. Because the sequence of two motions can be switched in the correlation calculation, the correlation maps given in the figure are symmetric matrices.}
        \label{fig:3.5}
\end{figure*}

\paragraph{Between-class Analysis}
This work aims to propose a machine learning-based multi-class classification method and reveal the relationship between the classification accuracy and the grand average MRCP of each motion. The confusion matrices and the correlation of the grand average MRCPs of motion pairs are compared. Fig. \ref{fig:3.4} shows the confusion matrices classified by the proposed mFBTRCA method. The confusion matrices are calculated using the 'confusionmat' function in MATLAB and are then normalized by dividing by the summed-up values in each row. Fig. \ref{fig:3.5} shows the two-dimensional Pearson correlation between the grand average MRCPs of motion pairs, which is calculated using the 'corr2' function in MATLAB. The grand average MRCPs used in the calculation of the Pearson correlation are spatial-filtered using the spatial filter $\boldsymbol{W}\in \mathbb{R}^{C\times P}$, and thus have the size $\mathbb{R}^{T\times P}$.

In Fig. \ref{fig:3.5a}, it can be seen that the grand average MRCP of \textit{elbow flexion} is highly correlated with \textit{elbow extension}. The EEG signals of both motions are difficult to discriminate with the classifier, as shown in Fig. \ref{fig:3.4a}. The \textit{resting} state has a significantly different grand average MRCP from the other movement states, and thus the classification between the $resting$ state and one of the motions outperforms the classification between two motions. Therefore, it is assumed that the classification of limb movements is highly correlated with the relationship between the grand average MRCPs. The analysis of dataset II also supports this assumption in Fig. \ref{fig:3.4b} and Fig. \ref{fig:3.5b}. For instance, the \textit{supination} and \textit{pronation} motions have a higher correlation between their grand average MRCPs, and the same is true for the \textit{palmar grasp} and \textit{lateral grasp} motions. Therefore, the classification between \textit{supination} and \textit{pronation} and between \textit{palmar grasp} and \textit{lateral grasp} achieves a worse performance than between other motion pairs, \textit{e.g.}, between \textit{supination} and \textit{lateral grasp} motions.

\section{Discussion}
\label{sec:discuss}

Non-invasive brain-computer interfaces can help humans to control external devices. The number of control commands is an important parameter in the applications of a brain-computer interface. In previous works, the control command was generated by analyzing passive-evoked brain activity (steady-state visual evoked potentials) or was limited by the number of limbs (motor imagery). This work focuses on the multi-class classification of upper limb movement, and aims at providing more active-evoked control commands in the brain-computer interface along with revealing the relationship between the classification performance and the grand average MRCP, which clarifies why certain pairs of upper limb movements cannot be accurately classified using EEG signals, \textit{e.g.}, \textit{elbow flexion} and \textit{elbow extension}.

The mFBTRCA method is developed by extending our previous proposed bFBTRCA method, which is a binary classification model. However, there is a drawback involved when migrating from binary to multi-class classification for the bFBTRCA method. The spatial filter in bFBTRCA is obtained by concatenating the eigenvectors of two classes. In the multi-class classification, the number of classes increases such that the number of eigenvectors in the spatial filters also increase. However, the number of eigenvectors in the spatial filter should be smaller than the number of channels, as otherwise the spatial filtering will increase the dimension along the channel axis of the EEG signals while keeping the rank unchanged; in other words, it would introduce useless information into the EEG signals.

The bFBTRCA method consists of three modules: spatial filtering, similarity measuring and filter bank selection. In the migration from the bFBTRCA method to the mFBTRCA method, it is assumed that spatial filtering plays the role of noise rejection, while the similarity measuring determines the classes of the EEG signals. This assumption is based on the processing of the grand average MRCP. 
The grand average MRCP of a class is obtained by averaging the EEG signals of that class. 
In MRCP signals, the class of the motions can be discriminated by comparing the differences of the grand average MRCPs. The purpose of averaging EEG signals is to remove random noise from the original signals. Therefore, the process of using the grand average MRCP to discern between different motions has two steps, (1) removing irrelevant noise from EEG signals and (2) discriminating classes by comparing the grand average MRCPs. These two steps are spatial filtering and similarity measuring, respectively.

The bFBTRCA method is migrated to mFBTRCA with this assumption. The optimization has two steps: (1) remove the steps related to the class information in the spatial filtering, and (2) reduce the similarity of the grand average MRCPs. The first step involves calculating the spatial filter with the eigenvectors given in Equation \ref{eqn:15}, and the second step involves removing the mean of all grand average MRCPs from the spatial-filtered EEG signals in Equation \ref{eqn:16}.

The mFBTRCA method achieves an equivalent classification performance to the bFBTRCA method in the binary classification task, as shown in Fig. \ref{fig:3.1}. Because the binary classification between a movement state and the \textit{resting} state (\textit{e.g.}, \textit{elbow flexion} vs \textit{resting}) has a higher classification accuracy, we assume that the classification performance between the motion pairs of the movement states will not fluctuate significantly. mFBTRCA is compared to state-of-the-art methods in classifying pairs of movement states and the \textit{resting} state (3 classes). The three-class classification accuracies are close to those achieved in the binary classification, but with slight decreases. For instance, the classification accuracy achieved between \textit{elbow flexion}, \textit{pronation} and \textit{resting} is close to the one achieved between \textit{elbow flexion} and \textit{pronation}.

The state-of-the-art methods used in this work include machine learning-based and deep learning-based methods. The compared machine learning-based methods are geared towards the multi-class classification of limb movements, such as those made by the left/right hand, the foot or the tongue. The classification of these motions is based on certain brain activity, such as motor imagery. The deep learning technique is a universal solution to classification. Although the deep learning methods are not specific to the classification of limb movements, they have a better performance than the machine learning methods. The proposed mFBTRCA method is also based on machine learning. However, mFBTRCA takes advantage of the differences of grand average MRCPs of motions. These differences are the reason why different limb movements can be classified. As a result, mFBTRCA performs better than the deep learning methods in the multi-class classification.

Besides the state-of-the-art methods, baseline methods were designed based on deep learning to be used for comparison purposes. In the design of these, we used the same idea as when designing the mFBTRCA method, namely using spatial filtering (spatial) and similarity measuring (temporal). The CNN layers were used to optimize the spatial characteristic of EEG signals. The RNN layers were then used to extract the temporal features. Finally, the CNN layer followed by a fully connected layer was the classifier used to predict the classes of EEG signals. We compared the mFBTRCA to the state-of-the-art and baseline methods in the multi-class classification task (more than three classes). As given in Table \ref{tab:3}, mFBTRCA shows an improved performance over the other compared methods, including the baseline methods. 

Although the proposed mFBTRCA method improves on the performance of the state-of-the-art and baseline methods, it also has its bottleneck. The method classifies the EEG signals based on the differences in the grand average MRCPs of motions. Importantly, when the grand average MRCPs of motions are correlated or almost the same, mFBTRCA fails to classify these motions, such as \textit{elbow flexion} and \textit{elbow extension}. Furthermore, the movement onset localization is a problem for armless or paralyzed patients in the actual application of the brain-computer interface. Our future work will focus on fusing the proposed mFBTRCA method with the deep learning techniques and applying transfer learning to help with the localization of the movement onset.

\section{Conclusion}
\label{sec:conclude}

We propose the mFBTRCA method for the multi-class classification of upper limb movements. The proposed method is comparable to the bFBTRCA method in binary classification. In the multi-class classification task (3 classes and more), mFBTRCA has a better performance than the other compared methods, including SCNN and baseline methods based on deep learning. In the 7-class classification of dataset I, mFBTRCA improves on the performance of the best-compared method by 4.19\%. In the 5-class classification of dataset II, the improvement is 8.73\%. The mFBTRCA method is developed by comparing the correlation between grand average MRCPs, thus revealing the relationship between the classification performance and the grand average MRCP. This method is expected to be the baseline in future works for the multi-class classification task of upper limb movements. 

\bibliographystyle{unsrt}
\bibliography{main}

\end{document}